\documentclass[12pt,twoside,a4paper]{article}
\usepackage{qmc,psfig}

\def\be{\begin{equation}}
\def\ee{\end{equation}}
\def\bea{\begin{eqnarray}}
\def\eea{\end{eqnarray}}
\def\ii{{\rm i}}
\def\tr{{\rm Tr}}
\def\nn{\nonumber}
\def\ex{{\rm e}}
\def\bfx{{\bf x}}
\def\lsi{\raise0.3ex\hbox{$<$\kern-0.75em\raise-1.1ex\hbox{$\sim$}}}
\def\gsi{\raise0.3ex\hbox{$>$\kern-0.75em\raise-1.1ex\hbox{$\sim$}}}
\newcommand{\lsim}{\mathop{\lsi}}

\begin{document}

\title{QCD thermodynamics at finite density
}
\author{
Owe Philipsen
}
\instit{
CTP, Massachusetts Institute of Technology,\\
Cambridge, MA 02139, USA
}
\gdef\theauthor{O.\ Philipsen}
\gdef\thetitle{Towards QCD thermodynamics at finite density}

\maketitle

\begin{abstract}
Recent progress in extending finite temperature lattice QCD simulations
from the $T$ axis into the finite density plane is reviewed. 
The covered topics are
a calculation of the transition line by multidimensional reweighting,
screening mass calculations in dimensionally reduced QCD and
simulations at imaginary $\mu$.
\end{abstract}

\section{Introduction}

\begin{figure}[htb]
\vspace{9pt}
\centerline{\psfig{file=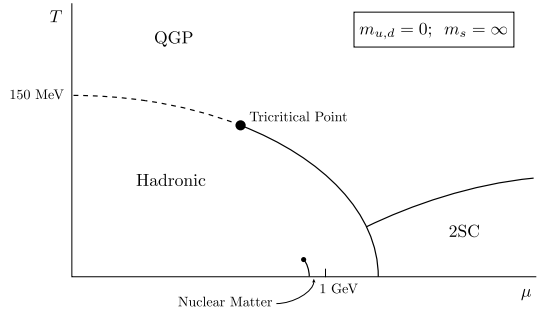,width=7cm}}
\caption[]{\label{pdiag}
Qualitative phase diagram for QCD with two massless flavours.
On the $T$ axis and along the dashed line there is a crossover,
solid lines mark first order transitions. The tricritical point
is expected to move closer to the $T$ axis for finite $m_s$.
From \cite{kr}.}\end{figure}

QCD at finite baryon density plays a role in nature in two rather different regimes:
It occurs in heavy ion collisions whose initial state has non-zero baryon number,
and whose subsequent plasma state is of high temperature and low density.
It is also believed to constitute the core of neutron stars, consisting
of cold and very dense matter.
These two situations correspond to the regions close to the axes in the
tentative QCD phase diagram Fig.~\ref{pdiag}.
Here the interest is in the former regime, whose understanding
is particularly pressing in view of experiments at SPS, LHC (CERN) and
RHIC (BNL).

In order to come to first principles predictions,
non-perturbative methods are required.
Unfortunately, lattice QCD has so far failed to be a viable tool for
the analysis because of the so-called ``sign-problem''. The QCD partition
function is given by
\be
Z=\int DU\, \det M(\mu)\ex^{-S_{YM}[U]},
\ee
where $S_{YM}$ is the pure gauge action and $\det M(\mu)$ 
denotes the fermion determinant.
For the gauge group SU(3) and $\mu\neq 0$, $\det M$ is complex and prohibits standard
Monte Carlo importance sampling, for which a manifestly positive measure in the functional
integral is required. While still no {\it ab initio} solution of the sign problem
is available, recent results suggest that at finite temperatures around 
the deconfinement
transition and for low densities the sign problem is milder and simulations 
are possible.
This contribution reviews some recent progress obtained by reweighting methods and
simulations of dimensionally reduced QCD. 
It also discusses what we can learn from QCD with
imaginary chemical potential where the integration measure is real and no reweighting
is necessary. For a more complete coverage and references to other 
approaches see e.g.~\cite{rev}.

\section{Reweighting methods}

The Glasgow method \cite{gla} evaluates the partition function by absorbing the
complex determinant into the observable, and doing the importance sampling with the
positive part of the measure,
\be
Z =\left\langle \frac{
\det M(\mu)}{\det M(\mu=0)}\right\rangle_{\mu=0}\;.
\ee
However, numerical results give for the onset of baryon density on the $\mu$ axis 
the unphysical value
$\mu_0=m_\pi/2$ instead of the expected $\mu_0=m_B/3$, 
which is also the pathological
behaviour of the quenched theory \cite{quen}.
The reason for this failure is the poor
overlap between the Monte Carlo ensemble of configurations at $\mu=0$ used to compute averages
by reweighting, and the full ensemble at $\mu\neq 0$.

This situation can be improved
by splitting the determinant in modulus and phase, $\det M = | \det M|\exp(\ii \phi)$,
and incorporating the modulus into the measure, while only
the phase is used for reweighting \cite{mod},
\be
Z =\left \langle 
\ex^{\ii\phi}\right\rangle_{|\det M|}\;.
\ee
The expectation value of the phase factor can be viewed as a ratio of two partition functions,
one of the full theory and one with only 
the positive modulus of the determinant in its measure. 
It thus behaves as $\exp (-\Delta F)$, where the exponent is the free energy difference and an
extensive quantity. Consequently,
$\langle \exp(\ii\phi)\rangle\sim\exp(-V)$,
whose measurement requires exponentially large statistics for realistic volumes.

\begin{figure}
\centerline{\psfig{file=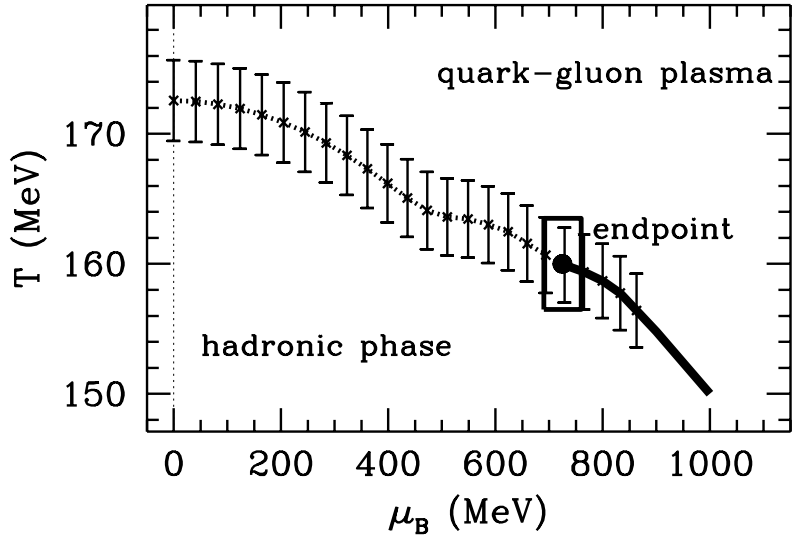,width=11cm}}
\caption[]{\label{fodor}
Phase diagram for 2+1 flavour QCD ($m_\pi$ is about four times its physical value, 
$m_s\sim 10 m_u$).
From \cite{fk}.}
\end{figure}

Recently considerable progress has been reported by means of a two-dimensional
reweighting method \cite{fk}, 
which in addition to $\mu$ also reweights in the lattice
gauge coupling $\beta$, rewriting the partition function as
\be
Z=\left\langle \frac{\ex^{-S_{YM}(\beta)}\det(M(\mu))}
{\ex^{-S_{YM}(\beta_0)}\det(M(\mu=0))}\right\rangle_{\mu=0,\beta_0}.
\ee
The simulation performed in \cite{fk} for 2+1 flavours of staggered quarks is concerned
with the critical line and the endpoint of the deconfinement phase transition. 
The simulation parameters
are tuned such as to stay on the (pseudo-)critical line 
defined by the location of Lee Yang zeros. 
While reweighting only in $\mu$ would inevitably
employ an ensemble away from criticality, the second parameter
can be used to keep the Monte Carlo ensemble fluctuating between the phases, as the
full ensemble certainly would.
This simulation 
constitutes the first numerical prediction of the $(\mu,T)$ phase diagram and
is displayed in Fig.~\ref{fodor}. 
However, it still has some caveats, and more work is required to confirm these results. 
The lattices are still rather coarse ($a\sim 0.3 fm)$ and the volumes
small. Since the sign problem becomes exponentially stronger with the volume,
it remains to be seen whether infinite volume and continuum limits can be taken with
this method.
More importantly, while the two-dimensional reweighting on the critical line
does sample both phases, there is no guarantee that it has a good overlap
with the full ensemble, which is still at a different point of parameter space. 
Hence it seems important to cross check these results by a different method.

\section{\label{imag} Imaginary chemical potential}

There are a number of suggestions to consider imaginary chemical potential, for which the
integration measure is positive and simulations are straightforward.
The connection to real chemical potential
is provided by the canonical partition function at fixed quark number $Q$, which is
related to the grand canonical partition function at fixed imaginary chemical potential
by \cite{rw,mr}
\be\label{can}
Z(T,Q) =
\int_{-\pi T}^{+\pi T} \frac{d \mu_I}{2\pi T}\, Z(T,\mu = \ii\mu_I)\,
e^{- i \mu_I Q/T}.
\ee
One proposal is to simulate $Z$ for various values of $\mu_I$, and then numerically do
the Fourier integration \cite{aw}. This becomes more and more difficult, however, 
for large $Q$, and extrapolation to the thermodynamic limit seems questionable.
The method has been tested in the two-dimensional Hubbard model \cite{aw} but not for QCD.

Another proposal, to be further pursued in the following, is
to use analyticity of the partition function to continue 
expectation values computed
with $Z(\ii\mu,T)$.
Not too far from the temperature axis (and away from a phase transition),
physical quantities are analytic functions of $\mu/T$ having a Taylor series.
One can then fit the coefficients of the truncated series to results obtained at
imaginary $\mu$ and analytically continue the series to real $\mu$.
This has been explored for the chiral condensate in the strong coupling limit
\cite{lom}. In the following it will be tested non-perturbatively in the 
deconfined phase in the framework of an effective theory.

\section{Dimensionally reduced QCD}

For temperatures larger than a few times the deconfinement temperature $T_c$,
the static (equilibrium) physics of QCD can be accurately described by an effective theory,
which indeed permits
simulations with non-vanishing real quark chemical potential 
$\mu/T\lsim 4.0$, where the sign problem sets in \cite{hlp1}. 
It permits
to study any number of quarks with small or zero mass. The domain of
parameter space where the approach is applicable contains the phenomenologically relevant
region in which heavy
ion collisions are operating. At and above SPS energies, densities in heavy ion collisions
are estimated to be $\mu_B/T\lsim 4.0$ \cite{exp}, i.e.~a quark chemical potential
$\mu/T\lsim 1.3$, which is well within the range where simulations are feasible.

Equilibrium physics is
described by euclidean time averages of gauge invariant operators,
$\bar{A}({\bf x})=T \int_0^{1/T} d\tau \,A(\bf{x},-\ii\tau)$.
Their spatial correlation functions
\be
C(|{\bf x}|)=\langle \bar{A}({\bf x})\bar{A}(0)\rangle_c\sim \ex^{-M|\bfx|},
\ee
decay exponentially with distance. The
``screening masses'' $M$ correspond to
the inverse length scale over which the equilibrated medium is sensitive
to the insertion of a static source carrying the quantum numbers of $A$.

For length scales larger than the inverse temperature, 
$|\bfx|\sim 1/gT\gg
1/T$, the time integration range
becomes very small and the problem effectively three-dimensional.
The calculation of the
correlation function $C(|\bf{x}|)$ can then be factorized:
the time averaging may be performed perturbatively
by expanding in powers of the ratio of scales $gT/T\sim g$,
which amounts to integrating out all modes with
momenta $\sim T$ and larger, i.e.~all non-zero Matsubara modes and 
in particular the fermions.
This procedure is known as dimensional reduction \cite{dr}. 
It is in the spirit of a
Wilsonian renormalization group approach, where an effective action for coarser scales
is derived by averaging over the smaller scales. 
The remaining correlation function of 3d fields is then to be evaluated
with a 3d purely bosonic effective action, which describes the physics of the
modes $\sim gT$ and softer.
Without fermions and one dimension less, accurate
infinite volume and continuum extrapolations of simulations are feasible.

The effective theory emerging from hot QCD by dimensional reduction
is the SU(3) adjoint Higgs model \cite{dr,rold,ad} with the action
\be \label{actc}
        S = \int d^{3}x \left\{ \frac{1}{2} \tr(F_{ij}F_{ij})
        +\tr(D_{i}A_0)^2 
+m_3^2 \tr(A_0^2) +\lambda_3(\tr(A_0^2)^{2} \right\} .
\ee
As 4d euclidean time has been integrated over,
$A_0$ now appears as a scalar in the adjoint representation.
The associated Hamiltonian respects
SO(2) planar rotations, two-dimensional parity $P$,
charge conjugation $C$ and $A_0$-reflections $R$, and screening masses can be 
classified by $J^{PC}_R$.

The parameters $\{g_3^2,m_3,\lambda_3\}$ of the effective theory are
via dimensional reduction functions of the 4d gauge
coupling $g^2$, the number of colours $N$ and flavours $N_f$, the fermion masses
and the temperature $T$. Note that $m_3\sim gT$ is just the leading order Debye mass.
In all of the following fermion masses are assumed to be zero,
but in principle any other values may be considered as well.

\begin{figure}
\leavevmode
\psfig{file=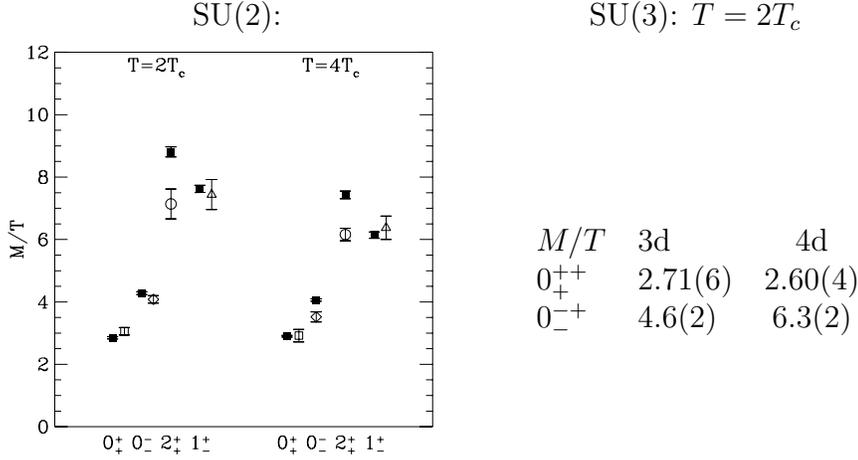,width=6cm}
\put(-100,170){SU(2):}
\put(50,170){SU(3): $T=2T_c$}
\put(20,70){
\parbox{6cm}{
\begin{tabular}{llc}
$M/T$ & 3d & 4d\\
$0^{++}_+$ & 2.71(6) & 2.60(4)\\
$0^{-+}_-$ & 4.6(2)  & 6.3(2)
\end{tabular}} }
\caption[]{\label{spec} Comparison of screening masses $J^P_R$ ($C=+$) in hot SU(2)
pure gauge theory as determined in 4d (empty symbols) and 3d (full symbols)
effective theory simulations. From \cite{hp}. Numbers for SU(3) from \cite{hlp1,dg}.}
\end{figure}

With the simulation part being rather accurate,
the main error in the correlation functions is due to the reduction step,
which has been performed to two loops \cite{ad}.
Fig.~\ref{spec} compares the results for hot pure gauge theory as obtained
in the 4d lattice theory \cite{dg} with those from the effective theory \cite{hp,hlp1}. 
Note that
the effective theory is only valid up to its cut-off $M/T\sim 2\pi$ and above this
level disagreement is to be expected.
For the case of SU(3) one finds
again quantitative agreement in the largest correlation length but about 20\%
deviation in the next shorter one.
Good agreement between the 4d and the 3d effective theory is also 
found with other observables,
like the static potential \cite{rold}, the spatial string tension \cite{bali}, 
the Debye mass \cite{debye}
and gauge-fixed propagators \cite{kar}. Dimensional reduction also works to high precision
from 3d finite $T$ to 2d \cite{bial}.
Thus, dimensionally reduced pure gauge theory gives a reasonable description
of the largest correlation lengths in the system down to temperatures as low
as $\sim 2T_c$.

\subsection{Finite density}

The main advantage of the effective theory is that fermions, 
having always non-zero Matsubara modes, are treated analytically. Any change in
the number of fermion species or their masses is encoded in the parameters of the
effective theory and only shifts the values for the screening masses in Fig.~\ref{spec}.
At $\sim2-3 T_c$, the
fermionic modes begin to feel the chiral phase transition and thus become very light through
non-perturbative effects. At some point this effect will be so large that they constitute
the lightest degrees of freedom and may no longer be integrated out,
as demonstrated in a 4d simulation with $N_f=4$ light fermions \cite{gg}.

When a chemical potential is switched on, its leading effect is to
generate one extra term in the action Eq.~(\ref{actc}) and to change
the Debye mass,
\bea \label{change}
S &\to& S + \ii z\int d^3x\,\tr A_0^3, \quad z=\frac{\mu}{T}\frac{N_f}{3\pi^2},\nn\\
m_3&\to &m_3\left[1+\Bigl( \frac{\mu}{\pi T}\Bigr)^2 \frac{3 N_f}{2 N + N_f}
\right]\,.
\eea
The extra term $S_z$ is odd under $R,C$, and hence the action no longer respects these
symmetries, while parity is left intact. 
Consequently, screening states now are labelled by $J^P$.

\begin{figure}[tb]%
\centerline{\psfig{file=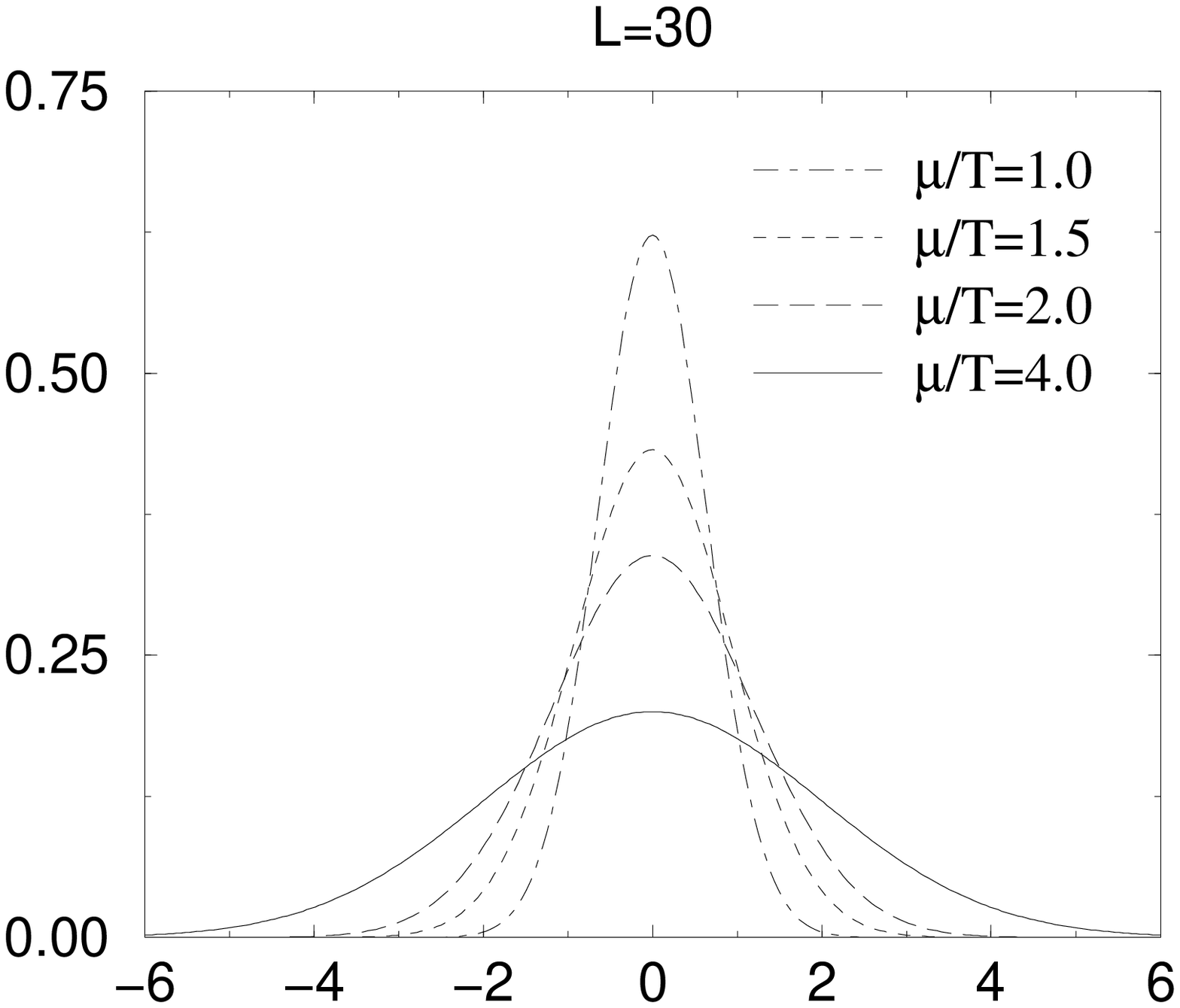,height=6cm}
\psfig{file=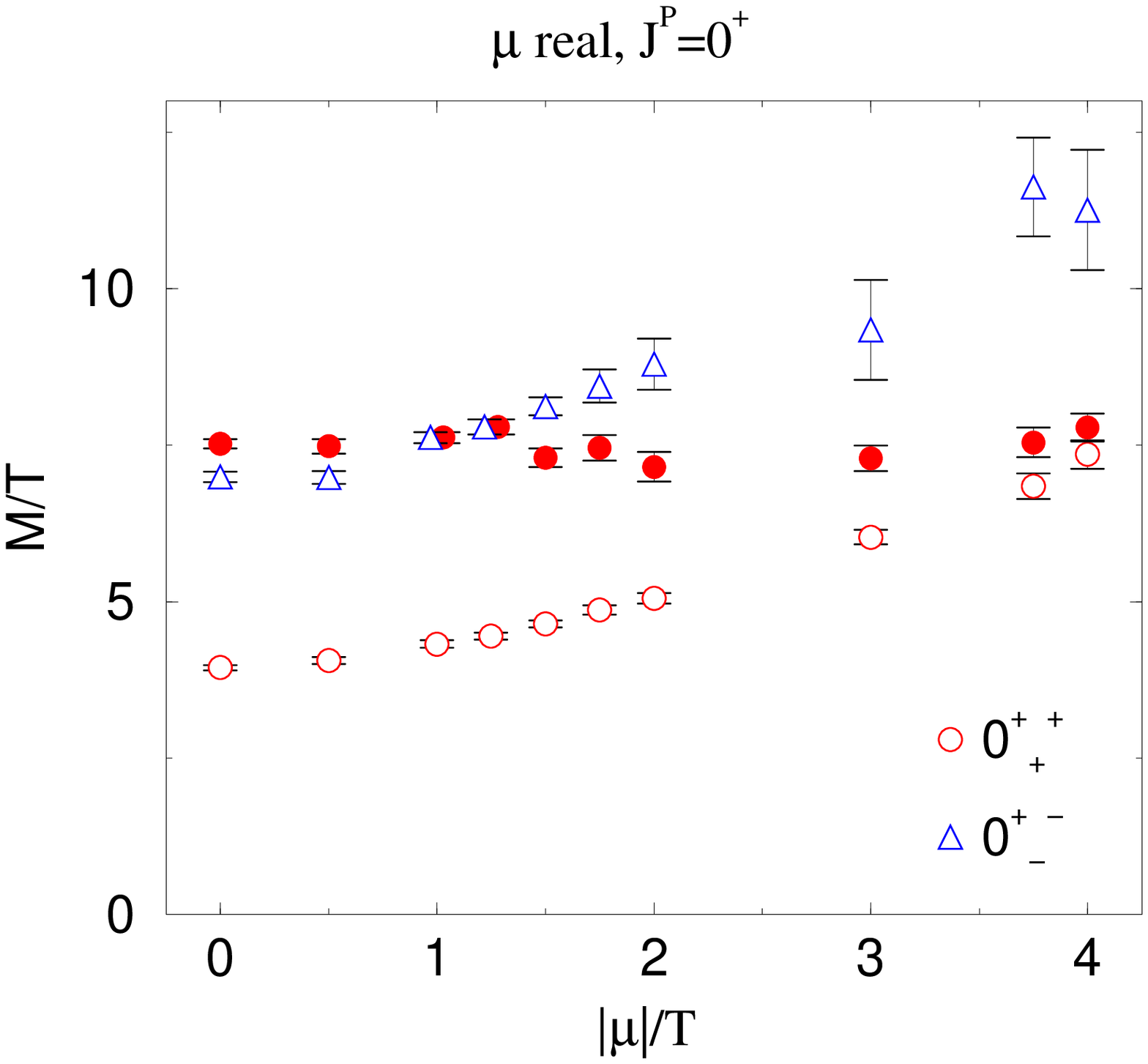,height=6cm}}
\caption[]{\label{dist} Results for $N_f=2$ and $T=2T_c$. 
Left: Distribution of the reweighting factor ($zS_z$) 
for fixed $L$.
Right: Screening masses in the channel $0^+$  $T=2T_c$.
`Scalar'' states ($\tr A_0^2$, empty symbols)
depend on $\mu/T$, while ``gluonic'' states ($\tr F_{12}^2$, full symbols )
are practically independent of it. From \cite{hlp1}.
}
\end{figure}

The effective action is complex and still
has a sign problem. Expectation values have to be computed
by reweighting with the complex piece of the action
\be
\langle {\cal O}\rangle =
\frac{\langle {\cal O} \ex^{-\ii zS_z}\rangle_0}{\langle\cos(zS_z)\rangle_0}\,,
\ee
where the subscript denotes averaging with $\mu=0$.
Cancelling contributions to the expectation value occur whenever
$(zS_z)\gg 1$. 
Fig.~\ref{dist}
shows some distributions of $(zS_z)$ as obtained by Monte Carlo for
chemical potentials. 
As long as $\mu\lsim 4T$,
the distribution is well contained within $[-\pi,\pi]$ for volumes large enough
so that the masses attain their infinite volume values without any cancellations
\cite{hlp1}. 
For even larger values
of the chemical potential the sign problem sets in and statistical errors explode.

The results for the lowest screening masses are also displayed in Fig.~\ref{dist}. 
The different qualitative behaviour of 3d gluonic and
scalar states observed generally in scalar gauge models \cite{tpw,hp,hlp1} leads
to a level crossing at $\mu/T\sim 4.0$ and hence 
to a change in the nature of the ground state excitation.
Hence there is interesting structure in the phase diagram even away from
a phase transition,
implying that the longest correlation length in the thermal system does
not get arbitrarily short with increasing density, but stays at
a constant level. 

\subsection{Imaginary vs. real chemical potential}

\begin{figure}[tbh]%
\centerline{\psfig{file=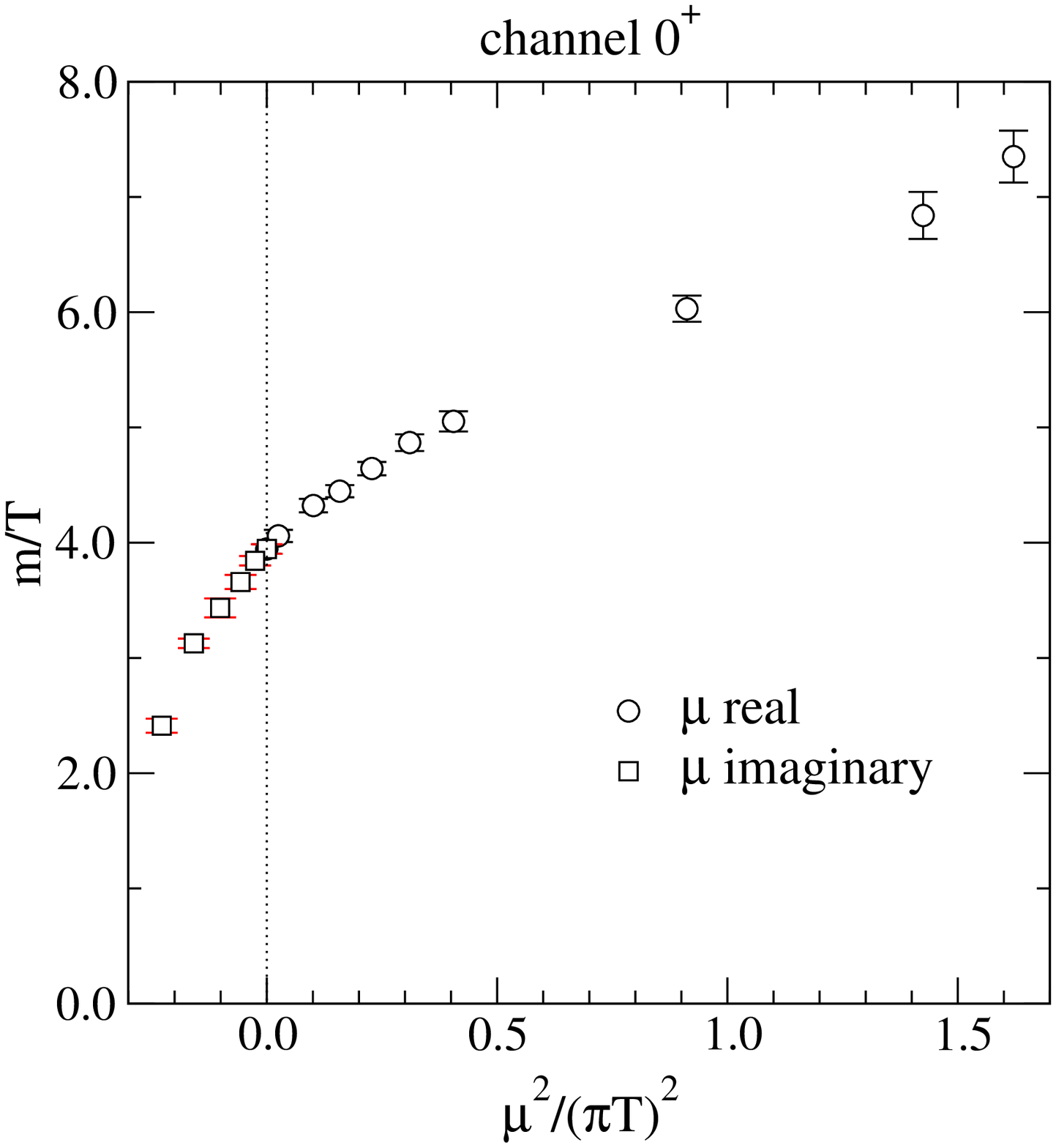,width=6cm}
\psfig{file=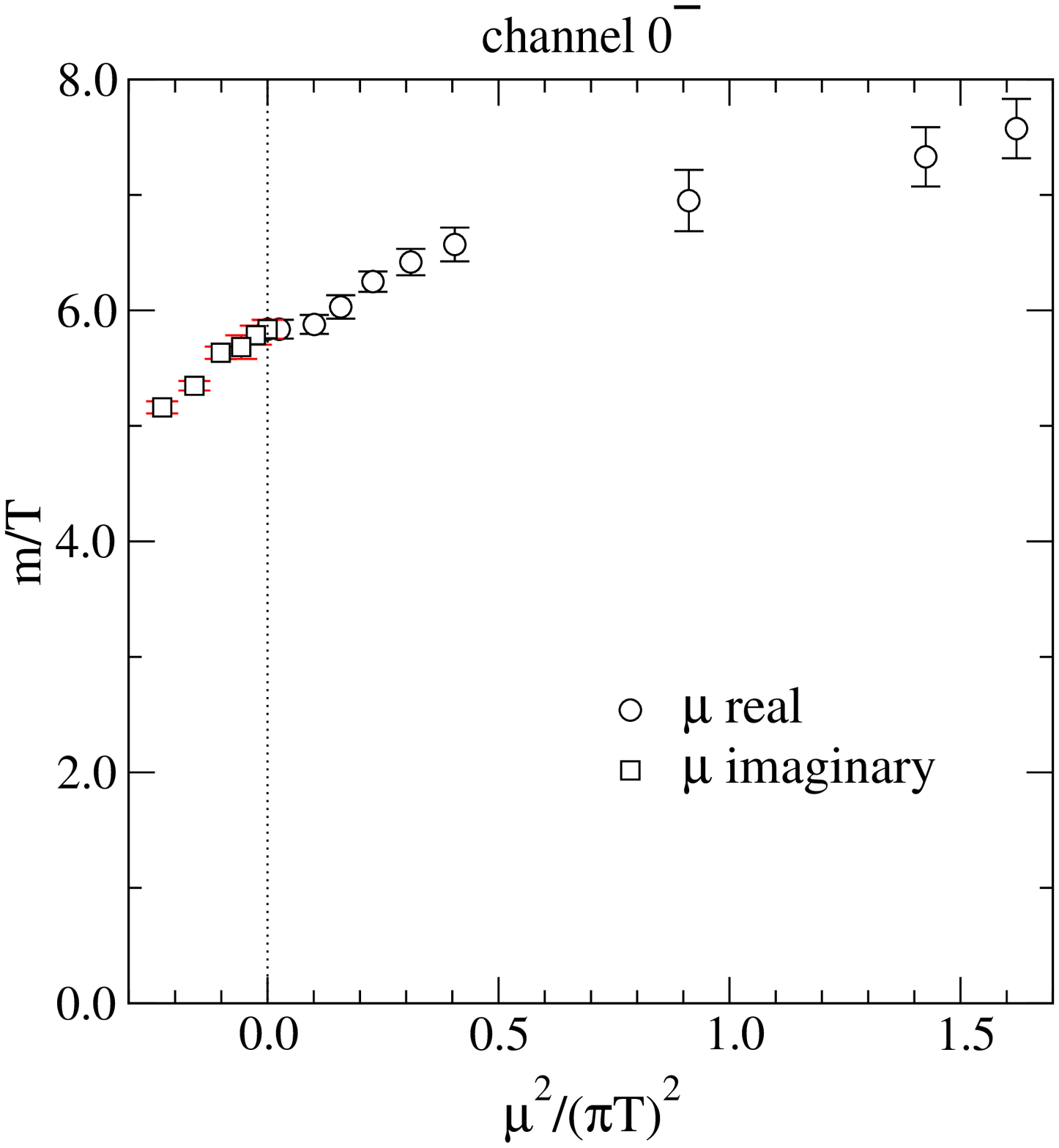,width=6cm}}
\caption[]{\label{cont} The lowest screening masses
at imaginary and real $\mu$, $T=2T_c$. From \cite{hlp2}.
}
\end{figure}
With an effective theory permitting simulation of real and imaginary chemical potential 
at hand, one may now
return to the suggestion of Sec.~\ref{imag} and study the feasibility
of analytic continuation of observables \cite{hlp2}. 
Away from phase transitions, the screening masses
are analytic in $\mu/T$, as there are no massless modes in the theory. Moreover,
since a change $\mu\rightarrow -\mu$ can be compensated for by a field
redefinition $A_0 \rightarrow -A_0$ in Eq.~({\ref{actc}), all physical
observables must be even under this operation. In the 4d
theory the same statement follows from compensating $\mu\to -\mu$ by a
C (or CP) operation. For small $\mu/T$,
the screening masses may thus be written
as
\be 
\frac{M}{T}  =  c_0
+ c_1 \biggl( \frac{\mu}{\pi T} \biggr)^2
+ c_2 \biggl( \frac{\mu}{\pi T} \biggr)^4\nn\\
+ {\cal O} \left( \frac{\mu}{\pi T} \right)^6 .
\label{ansatz}
\ee
In the range where this ansatz fits the data of real and imaginary $\mu$, 
the possibility of analytic continuation is easily checked by examining whether
the $c_i$ agree between the two data sets.

Fig.~\ref{cont} shows the lowest states in the system for real and imaginary potential.
Note that in the imaginary $\mu$ case, data are only available up to $|\mu|/T\sim 1.5$,
and hence $c_2$ is not well constrained.
The reason is that imaginary chemical potentials favour a $Z_N$-broken
minimum over the symmetric one in the 4d effective $A_0$ potential, 
once $|\mu|/T\geq\pi/3$.
Thus a phase transition occurs and analyticity is lost \cite{rw}.

In the region up to this critical value, however, 
one observes firstly that the coefficients $c_{0,1}$ are sufficient 
to fit the data, and secondly that they are completely compatible
between the real and imaginary $\mu$ data sets \cite{hlp2}.
Having only two coefficients is equivalent to susceptibility measurements
at $\mu=0$. The first conclusion then endorses work based on that 
latter approach \cite{susfree}. The second conclusion  
encourages us to consider 4d QCD at imaginary $\mu$ and study the vicinity
of the critical line.

\section{QCD at imaginary chemical potential}

From the fact that screening masses decrease with imaginary $\mu$ it follows that
$T_c(Im(\mu))$ is an increasing function, leading to
Fig.~\ref{imdiag} (left) as the complex generalization of the phase 
diagram Fig.~\ref{pdiag}.
\vspace*{-0.5cm}
\begin{figure}[tbh]%
\leavevmode
\hspace*{0.5cm}
\psfig{file=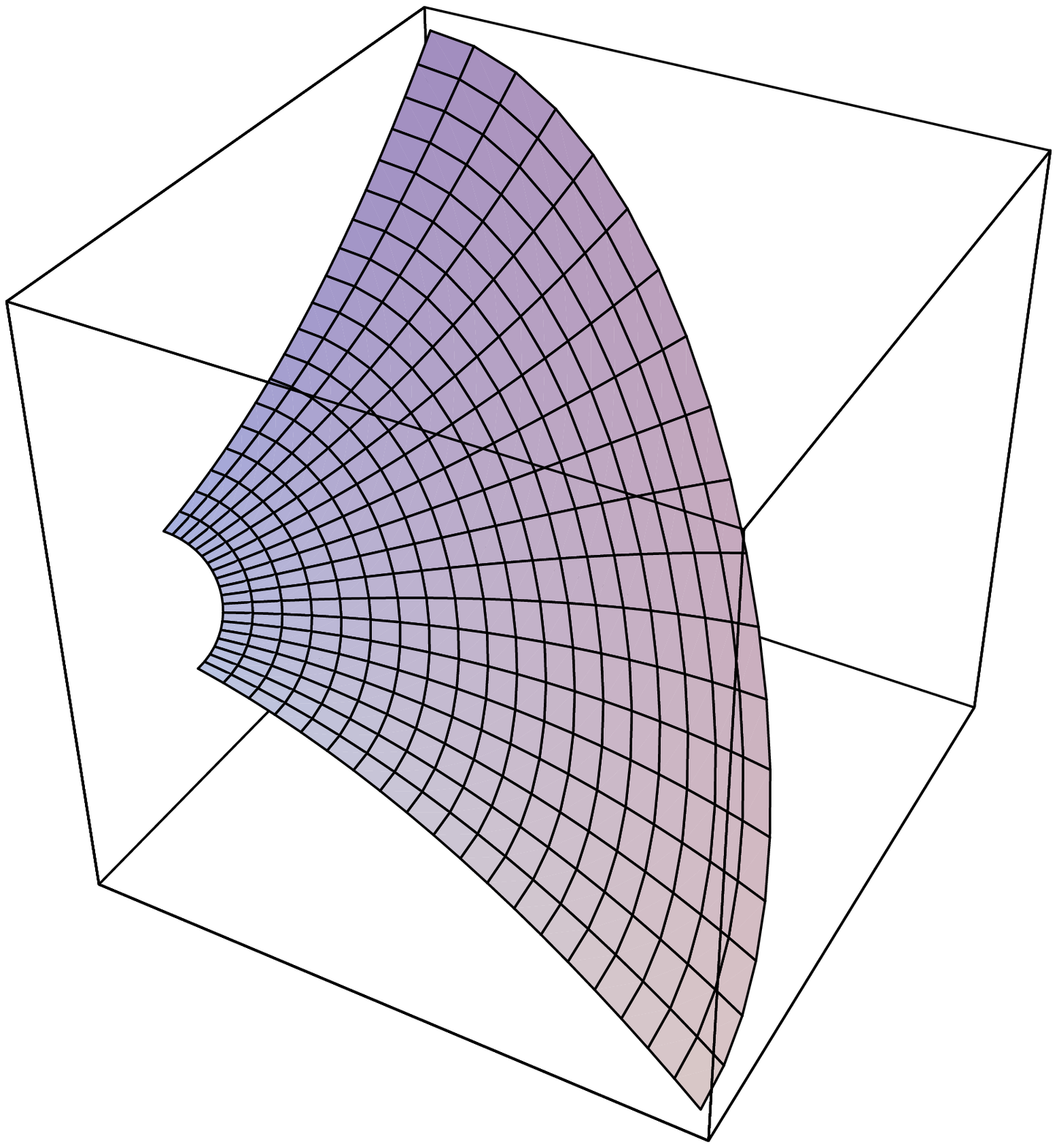,height=7cm}
\put(-125,25){\small $Re(\mu)$}
\put(-130,165){\small $Im(\mu)$}
\put(-160,100){\small $T$}
\psfig{file=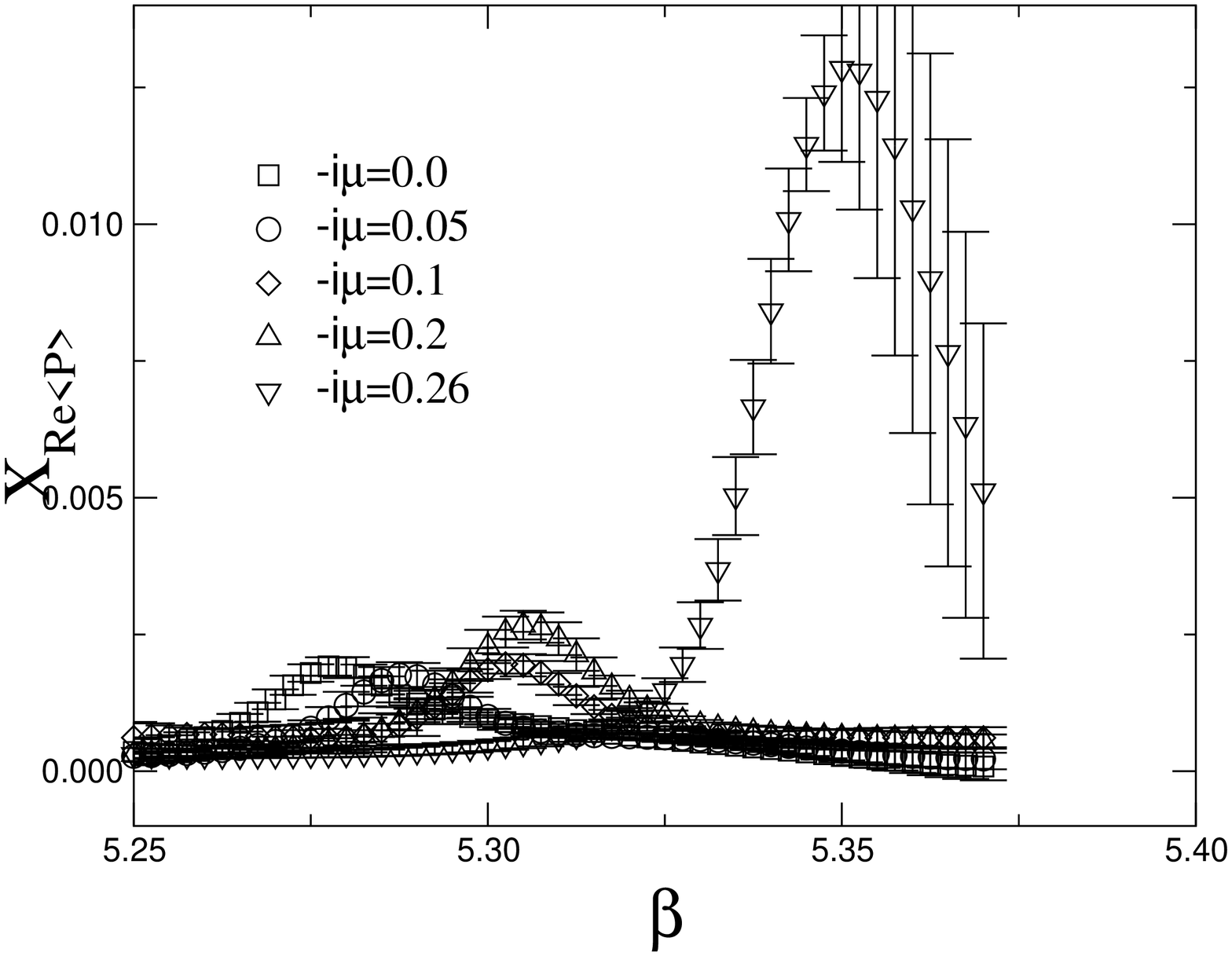,height=6cm,angle=-90}
\caption[]{\label{imdiag} Left: Qualitative phase diagram for complex $\mu$.
Right: Susceptibility of the Polyakov loop for imaginary $\mu$. 
$T$ increases with $\beta$.
}
\end{figure}
First results of simulating two-flavour QCD with staggered fermions at
imaginary $\mu$ \cite{us} fully confirm this picture, as shown in
Fig.~\ref{imdiag} (right). The peak of the susceptibility of a gauge invariant
operator occurs at a (pseudo-)critical point. The figure clearly
shows that the critical coupling $\beta_c$ (and with it $T_c$)
is growing with $|\mu|$. Furthermore, on finite volumes susceptibilities
are analytic funtions. This means that $T_c(\mu)$ itself is 
analytic and it might be possible to continue it from imaginary to real $\mu$.
This question is currently under investigation.

\section{Conclusions}

Recent developments have shown that at high temperature, 
despite the sign problem, lattice
simulations can be extended away from the temperature axis 
into the finite density plane. Considering
dimensionally reduced QCD gives a detailed picture
of screening masses in the deconfined phase, covering
the experimentally relevant parameter space. 
Furthermore, it establishes that observables depend only weakly on 
$\mu$ and hence susceptibility measurements as well as analytic
continuation from imaginary to real $\mu$ are possible.
Multidimensional reweighting
has even produced a result for the critical line of the deconfinement
transition and its endpoint. Because it uses reweighting, however, an
independent check would be desirable. 


\end{document}